\documentclass[aps,prc,preprint,superscriptaddress,showpacs]{revtex4}
\usepackage{graphicx}
\usepackage{dcolumn}
\usepackage{bm}

\newcommand{\rgn}{($\gamma$,n)}
\newcommand{\rng}{(n,$\gamma$)}

\newcommand{\wovi}{$^{186}$W}
\newcommand{\wov}{$^{185}$W}
\newcommand{\spro}{$s$-process}

\begin{document}

\title{
$\mathbf{s}$-process branching at \wov\ revised
}

\author{P.\ Mohr}
\email{WidmaierMohr@compuserve.de}
\affiliation{
  Institut f\"ur Kernphysik, Technische Universit\"at Darmstadt,
  Schlossgartenstra{\ss}e 9, D-64289 Darmstadt, Germany
}
\affiliation{
  Strahlentherapie, Diakoniekrankenhaus Schw\"abisch Hall,
  D--74523 Schw\"abisch Hall, Germany
}
\author{T.\ Shizuma}
\affiliation{
  Advanced Photon Research Center, Japan Atomic Energy Research Institute, \\  
  Tokai, Ibaraki 319-1195, Japan  
}
\author{H.\ Ueda}
\author{S.\ Goko}
\author{A.\ Makinaga}
\author{K.Y.\ Hara}
\affiliation{
  Department of Physics, Konan University, \\
  8-9-1 Okamoto, Higashinada, Kobe 658-8501, Japan
}
\author{T.\ Hayakawa}
\affiliation{
  Advanced Photon Research Center, Japan Atomic Energy Research Institute, \\ 
  Tokai, Ibaraki 319-1195, Japan  
}
\affiliation{
  National Astronomical Observatory, Osawa 2-21-1, Mitaka, 
  Tokyo 181-8588, Japan
}
\author{Y.-W.\ Lui}
\affiliation{
  Cyclotron Institute, Texas A \& M University, \\
  College Station, Texas 77843, USA
}
\author{H.\ Ohgaki}
\affiliation{ 
  Institute of Advanced Energy, Kyoto University,\\
  Gokanosho, Uji, Kyoto 611-0011, Japan
}
\author{H.\ Utsunomiya}
\affiliation{
  Department of Physics, Konan University, \\
  8-9-1 Okamoto, Higashinada, Kobe 658-8501, Japan
}
\date{\today}

\begin{abstract}
The neutron capture cross section of the unstable \spro\ branching
nucleus \wov\ has been derived from experimental data of the inverse
\wovi \rgn \wov\ photodisintegration taken with monochromatic
photon beams from laser Compton scattering. The result of
$\sigma = 553 \pm 60$\,mb at $kT = 30\,$keV leads to a relatively high
effective neutron density in the classical \spro\ of $N_{\rm{n}} =
4.7 \times 10^8\,{\rm{cm}}^{-3}$. A realistic model for the \spro\ in
thermally pulsing AGB stars overestimates the abundance of $^{186}$Os
significantly because of the relatively small neutron capture cross
section of \wov .
\end{abstract}


\maketitle

About half of the nuclei heavier than iron have been synthesized by a
series of neutron capture reactions and subsequent $\beta$-decays in
the so-called astrophysical \spro . This process is called slow
because the neutron capture rate is smaller than the $\beta$-decay rate
for most unstable nuclei. However, there is a number of relatively
long-living unstable nuclei with typical half-lives of at least
several weeks where neutron capture can compete with the
$\beta$-decay. Such nuclei are called branching points of the \spro\
because nucleosynthesis proceeds partly on a neutron-rich branch and
partly on a neutron-deficient branch. The analysis of branching ratios
allows to determine the effective neutron density $N_{\rm{n}}$ during
the \spro\ in a simple model, the so-called classical \spro\
\cite{Ward76}. Alternatively, branching points provide a stringent test
for realistic \spro\ models which describe the dynamics of thermally
pulsing asymptotic giant branch (AGB) stars in combination with the
corresponding neutron production and nucleosynthesis by
neutron-induced reactions \cite{Gal98,Arla99,Bus01}.

Despite the experimental progress with high-intensity neutron sources
and tiny amounts of target material, it still remains extremely
difficult to measure the neutron capture cross section of relatively
short-living nuclei like \wov\ with half-lives of $t_{1/2} \ll 1$\,y.
This neutron capture cross section for \wov\ ($t_{1/2} = 75.1$\,d) may
be derived from the inverse \wovi \rgn \wov\ photodisintegration with
help of theoretical models.

A first experiment at low energies was performed by Sonnabend
{\it et al.}\ \cite{Sonn03} using bremsstrahlung photons and the
photoactivation method. Because of the broad bremsstrahlung spectrum
it was not possible to measure the energy dependence of the \rgn\
reaction in that experiment. Additionally, the result had significant
systematic uncertainties of about 15\,\% because of the uncertainties
of the shape of the bremsstrahlung spectrum close to its endpoint
energy. Therefore we remeasured the \rgn\ cross section of \wovi\ using
a tunable monochromatic photon source from laser Compton scattering
(so-called Laser Compton Scattering photons, LCS).

The experiment was performed at AIST (Tsukuba, Japan). Photons from a
frequency-doubled Nd:YLF Q-switch laser at a wavelength of $\lambda =
526$\ nm were $180^\circ$ scattered from a relativistic electron beam
in the storage ring TERAS. The electron energy was varied from 460 to
588\,MeV which allowed to produce photons with maximum energies from
7.4 to 12.2\,MeV. The initial electron current of about 200\,mA in the
storage ring combined with the high laser power of 40\,W leads to a
typical photon intensity of about $10^4$/s after collimated into a 2
mm (in diameter) spot at target position which is located roughly 8\,m
from the interaction area of the laser photons and electrons. The
number of LCS photons decreases during the experiment with the
decreasing electron beam current; therefore the electron storage ring
was refilled twice a day. Further details on the LCS photon setup at
AIST and its application to photonuclear astrophysics can be found in
\cite{Ohga91,Toyo2000,Uts01,Uts03}.

The target consisted of 1246\,mg metallic tungsten powder highly
enriched in \wovi\ to 99.79\,\%. The powder was pressed to a small
self-supporting tablet with a diameter of 8\,mm. The tablet was
mounted into a thin holder made of pure aluminum which does not emit
neutrons below its high neutron separation energy of $S_{\rm{n}} =
13.1$\,MeV. The neutrons from the \wovi \rgn \wov\ photodisintegration
were detected using an improved neutron detector that consists of 16
individual $^3$He counters embedded in two rings in a polyethylene
moderator. The so-called ring ratio between the count rates of the
inner and outer rings depends on the neutron energy, and hence the
ring ratio can be used to estimate the neutron energy. In this
experiment the ring ratio varied between 2.5 and 4.4 leading to
average neutron energies of about 1.2\,MeV at highest photon energies
and of about 0.3\,MeV at lower neutron energies for the measurements
close above the threshold. Further experimental details are given in
\cite{Uts03}. The efficiency of the neutron detector is given in
Fig. 2 of \cite{Kobe}; the efficiency was measured at the average
neutron energy 2.14 MeV using a calibrated $^{252}$Cf source and the
energy dependence was determined by a MCNP simulation.


The number of neutrons $n_{\rm{exp}}$ emitted in the
photodisintegration experiment is directly related to the \rgn\ cross
section $\sigma(E_\gamma)$ for ideally monochromatic photons with
energy $E_\gamma$
\begin{equation}
n_{\rm{exp}} = N_\gamma \, \times \, N_{\rm{T}}\, \times \, h
\times \, \sigma(E_\gamma),
\label{eq:sigma}
\end{equation}
where $N_\gamma$ is the number of photons, $h$ is the correction
factor for a thick-target measurement, $ h = (1 - e^{- \mu t})/ \mu t$
with the target thickness $t$ and the attenuation coefficient of
target material $\mu$, and $N_{\rm{T}}$ is the number of target atoms
per area. For the realistic photon spectrum with a low-energy tail the
product $N_\gamma \times \sigma(E_\gamma)$ in Eq.~(\ref{eq:sigma}) has
to be replaced by the integral
\begin{equation}
N_\gamma \times \sigma(E_\gamma) \, \, \rightarrow \, \,
\int{n_\gamma(E_\gamma) \times \sigma(E_\gamma) \, dE}
\label{eq:spec}
\end{equation}
with the photon energy distribution $n_\gamma(E_\gamma)$.  In
Eq.~(\ref{eq:spec}), let us rewrite $\sigma(E_\gamma)$ in the Taylor
series,
\begin{equation}
\sigma(E_\gamma) = \sigma(E_{0}) + \sigma^{(1)}(E_{0})(E_\gamma - E_{0}) \,
+ \frac{1}{2} \sigma^{(2)}(E_{0})(E_\gamma - E_{0})^{2} \,
+ \frac{1}{6} \sigma^{(3)}(E_{0})(E_\gamma - E_{0})^{3} + \cdot \cdot \cdot,
\label{eq:taylor}
\end{equation}

\noindent
where $\sigma^{(i)} = d^{i} \sigma(E)/dE^{i}$.  When the average energy is
chosen for $E_{0}$, putting the Taylor series into Eq.~(\ref{eq:spec})
ends up with
\begin{equation}
\int{n_\gamma(E_\gamma) \times \sigma(E_\gamma) \, dE_\gamma} = \,
N_\gamma \{ \sigma(E_{0}) + s_{2}(E_{0}) + s_{3}(E_{0}) + \cdot \cdot \cdot \},
\label{eq:series}
\end{equation}

\noindent
where $s_{2}(E_{0}) = \frac{1}{2} \sigma^{(2)}(E_{0}) \, [
\bar{E_\gamma^{2}} - E_{0}^{2} ]$ and $s_{3}(W_{0}) = \frac{1}{6}
\sigma^{(3)}(E_{0}) [ \bar{E_\gamma^{3}} - 3 E_{0}
\bar{E_\gamma^{2}} + 2 E_{0}^{3} ]$ with $\bar{E_\gamma^{i}} =
\int{n_\gamma(E_\gamma) E_\gamma^{i}} d E_\gamma / N_\gamma$.  Note
that the first derivative term $\sigma^{(1)}$ explicitly vanishes.

Experimentally, the whole Taylor series in the parenthesis in
Eq.~(\ref{eq:series}) is obtained by using the numbers of neutrons,
target nuclei per unit area, and incident $\gamma$ rays.  In contrast,
the first term $\sigma(E_{0})$, which is the cross section at the
average $\gamma$ energy, is obtained provided that the $s_{2}$ and the
$s_{3}$ {\it etc.} are subtracted.  

We evaluated the higher-order terms which include $\sigma^{i}(E_{0})$
and $\bar{E_\gamma^{i}}$, where the energy dependence of the cross
section was derived by the best fit to the experimental quantity
corresponding to the whole Taylor series plotted at the average
$\gamma$ energy.  Details of the evaluation will be given in a
separate paper, including a more general discussion on the methodology
of deducing cross sections in a quasi-monochromatic $\gamma$-induced
reaction. The subtraction of the higher-order terms resulted in a few
\% increase in $\sigma(E_{0})$ in the energy region of astrophysical
relevance below 8.6 MeV and a decrease of 12 - 20 \% above 9 MeV.

\begin{figure}[hbt]
\includegraphics[ bb = 116 255 450 575, width = 75 mm, clip]{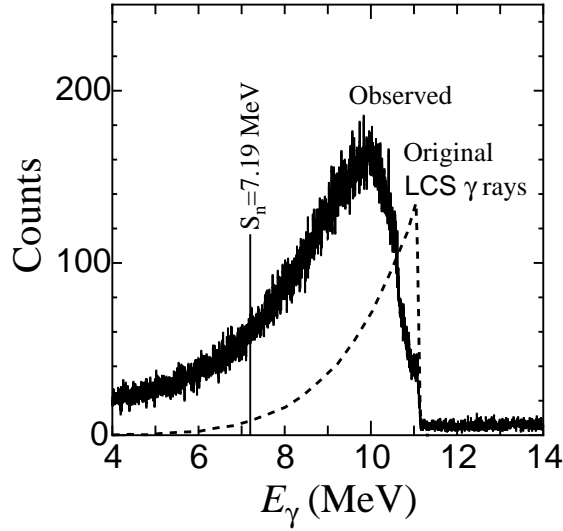}
\caption{
\label{fig:spec_ge}
Photon spectrum measured with a 120\,\% relative efficiency germanium
(HPGe) detector (full line) along an incident photon
spectrum (dashed line). For the analysis of the measured HPGe
spectrum to obtain the incident LCS photon energy spectrum, see
\cite{Uts03}. The neutron separation energy $S_{\rm{n}}$ of \wovi\ is
indicated by a vertical line.  
}
\end{figure}

A typical photon spectrum $n_\gamma(E_\gamma)$ is shown by the dashed
line in Fig.~\ref{fig:spec_ge}. The time variation of the photon
spectrum which is sensitive to the electron beam size at the collision
point was carefully investigated. The analysis of the photon spectrum
measured with the HPGe detector showed that the variation of the
electron beam current during individual measurements did not result in
a significant change in the beam size. Thus, both the average
$\gamma$ energy and the fraction of the photon spectrum above the
neutron threshold were determined with sufficient accuracy.

\begin{figure}[hbt]
\includegraphics[ bb = 115 285 480 605, width = 75 mm, clip]{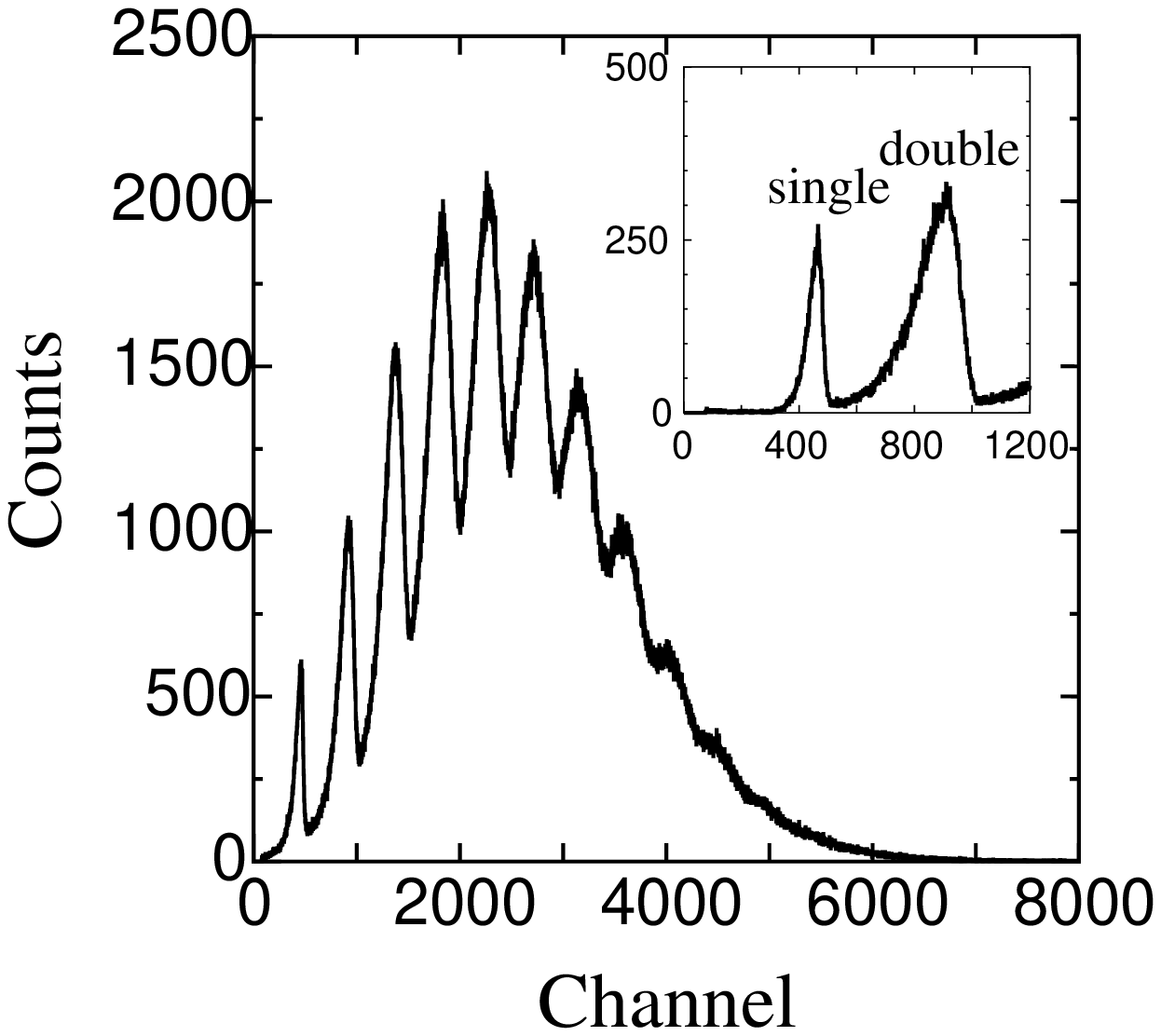}
\caption{
\label{fig:spec_sum}
Photon spectrum measured with a 8 x 12 inches NaI(Tl) detector. The number
of photons per laser pulse can be extracted from the average pulse output of
the summing crystal. In the shown spectrum the average photon number per
laser pulse is 6.2 leading to a photon intensity of $1.2 \times
10^4$/s. The inset shows a similar spectrum measured with reduced
laser power; here one finds mainly one or two LCS photons per laser
pulse. 
}
\end{figure}

The number of incoming photons was monitored during the
experiment using a large-volume 8 x 12 inches (diameter x length)
NaI(Tl) summing crystal. The pulse height in the sum spectrum is
proportional to the number of LCS photons which have the 2\,kHz repetition
rate of the laser. A typical spectrum of the NaI(Tl) summing
crystal is shown in Fig.~\ref{fig:spec_sum}.

\begin{figure}[hbt]
\includegraphics[ bb = 35 33 293 267, width = 75 mm, clip]{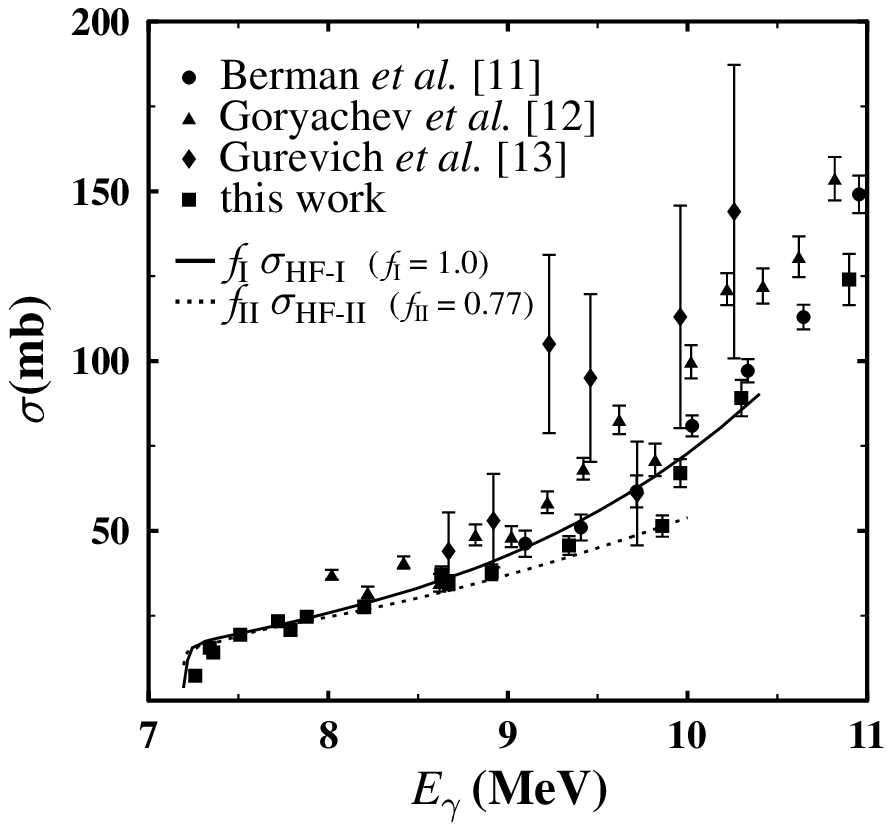}
\caption{
\label{fig:sigma}
Photodisintegration cross section of the reaction \wovi \rgn \wov .
The new data are shown as squares. Previous data of
\cite{Ber69,Gor78,Gur81}, shown as circles, triangles, and diamonds,
do not cover the astrophysically relevant energy region close above
the threshold at $S_{\rm{n}} = 7194$\,keV. The statistical model
predictions (as taken from Ref.~\cite{Sonn03}) have been scaled by
factors of $f_{I} = 1.0$ (full line) and $f_{II} = 0.77$ (dotted line)
to fit the new experimental data (see text).
}
\end{figure}

The measured photodisintegration cross section from the threshold of
\wovi\ at $S_{\rm{n}} = 7194$\,keV up to about 11\,MeV is shown in
Fig.~\ref{fig:sigma}. 
The systematic uncertainties are dominated by the efficiency of the
neutron detector (5\,\%) and the $\gamma$-ray flux (3\,\%).  At energies close to the threshold statistical uncertainties are comparable
to the above systematic uncertainties; at higher energies statistical
uncertainties are small. The error bars shown in Fig.~\ref{fig:sigma}
include both systematic and statistical uncertainties. Compared to the
previous experiment \cite{Sonn03}, the uncertainties have been reduced
significantly. More importantly, the energy dependence of the
photodisintegration cross section has been determined down to the
threshold. Earlier data at higher energies \cite{Ber69,Gor78,Gur81}
are in reasonable agreement with our new data but have larger
uncertainties especially at lower energies.

There is no direct way to derive the neutron capture cross section of
the \spro\ branching nucleus \wov\ from the photodisintegration cross
section of the \wovi \rgn \wov\ reaction. Following
\cite{Sonn03}, the theoretical prediction is adjusted to the
experimental data using a normalization factor
$f_{(\gamma,{\rm{n}})}$. The same normalization factor
$f_{({\rm{n}},\gamma)} = f_{(\gamma,{\rm{n}})}$ is used to scale the
theoretical prediction of the neutron capture cross section. This
procedure, its reliability and limitations are discussed in detail in
\cite{Sonn03}. Here we repeat briefly the basic idea. For the above
reactions the main ingredients for the statistical model predictions
are the photon strength function, the neutron-nucleus optical
potential, and the level densities. It has turned out that the most
sensitive ingredient is the electric dipole (E1) photon strength
function which is usually extrapolated from the GDR to lower
energies. Both $\sigma(\gamma,{\rm{n}})$ and $\sigma({\rm{n}},\gamma)$
are proportional to this strength function, and consequently
the above assumption $f_{({\rm{n}},\gamma)} = f_{(\gamma,{\rm{n}})}$
is justified if all other ingredients of the model are precisely
known. The assumption approximately remains valid for realistic cases
to within $10\,\% - 20\,\%$ because of the
uncertainties of the other ingredients. An obvious additional
requirement to the theoretical model is the correct prediction of the
energy dependence of the photodisintegration cross section which is
fulfilled for both calculations of \cite{Sonn03}, at least at energies
close above the threshold (see Fig.~\ref{fig:sigma}).

Two statistical model (or Hauser-Feshbach, HF) calculations with
different ingredients (called $I$ and $II$) were used in
\cite{Sonn03} to derive the neutron capture cross section from the
experimental photodisintegration data. Model $I$ predicts a neutron
capture cross section of $\sigma_{\rm{pre}} = 600$\,mb at $kT =
30$\,keV. Together with the scaling factor $f_{(\gamma,{\rm{n}}),I} =
1.0$ and the above assumption of $f_{({\rm{n}},\gamma)} =
f_{(\gamma,{\rm{n}})}$ one obtains the experimentally corrected cross
section of $\sigma_{\rm{exp}} = 600$\,mb. This value is already
Maxwellian averaged for a temperature $kT = 30$\,keV, and it includes
a minor correction for thermally excited states in \wov\ at such
temperatures. The corresponding values for model $II$ are
$\sigma_{\rm{pre}} = 657$\,mb, $f_{(\gamma,{\rm{n}}),II} = 0.77$, and
$\sigma_{\rm{exp}} = 506$\,mb. Averaging both values of
$\sigma_{\rm{exp}}$, the final result for the neutron capture cross
section of \wov\ is $\sigma = 553 \pm 60$\,mb. The uncertainty of this
value is dominated by theoretical uncertainties for the relation
between the \rgn\ and \rng\ reactions which can be estimated to be
47\,mb from the 
deviations of the two calculated values. The experimental
uncertainties of the present \rgn\ data are much smaller.

The new result is about 20\,\% lower than the previous result of
\cite{Sonn03}; taking into account the 15\,\% uncertainty of the
experimental data of \cite{Sonn03}, there is reasonable agreement
between the previous data and the new experimental results. The new
result is also slightly lower than the adopted value
($\sigma_{\rm{adopt}} = 703 \pm 113$\,mb) of a recent
compilation \cite{Bao00}; the value from the compilation is based on
several theoretical predictions \cite{Rau00,Hol76,Kae91}.

There are interesting astrophysical consequences of this new result
for the neutron capture cross section of \wov . The derived neutron
density $N_{\rm{n}}$ in the classical \spro\ scales inversely with the
neutron capture cross section of the analyzed branching nucleus. A
relatively high value of $N_{\rm{n}} = (4.7^{+1.4}_{-1.1}) \times
10^8\,{\rm{cm}}^{-3}$ is obtained from the new value of $\sigma =
553$\,mb. A realistic \spro\ model \cite{Gal98,Arla99,Bus01} describes
the \spro\ during thermally pulsing AGB stars. A cross section of
about 1000\,mb is required to reproduce the abundance of $^{186}$Os
which depends on the branching at \wov . The previous value of $\sigma
= 687$\,mb \cite{Sonn03} leads to an overproduction of $^{186}$Os of
20\,\%; taking into account the uncertainties of the solar osmium
abundance and of the $^{186}$Os neutron capture cross section (as
discussed in \cite{Sonn03}), the \spro\ model prediction corresponds
to an error at the 3\,$\sigma$ level. The even smaller cross section
of $\sigma = 553$\,mb of this work sharpens the discrepancy with the
otherwise successful model of the \spro . Hence the new data provide
further restrictions for realistic \spro\ models and may contribute to
improve such models.

\begin{acknowledgments}
We thank N.\ Pietralla for the borrowing of the enriched target. Discussions
with H.\ Beer, R.\ Gallino, F.\ K\"appeler, A.\ Mengoni, T.\ Rauscher,
and A.\ Zilges are gratefully acknowledged. This work was supported in
part by the Japan Private School Promotion Foundation and by the Japan
Society for the Promotion of Science.  
\end{acknowledgments}

\end{document}